\documentclass[traditabstract]{aa}
\usepackage{txfonts}
\usepackage{graphicx}
\usepackage{graphicx}
\usepackage{epsfig}
\usepackage{color}
\usepackage{epsf}
\usepackage{longtable,lscape}
\usepackage{rotating}
\usepackage{thumbpdf}
\usepackage{natbib}
\usepackage{color}

\begin{document}

\title{Distortion of the ultrahigh energy cosmic ray flux from
  rare transient sources in inhomogeneous extragalactic magnetic
  fields} \author{Sihem Kalli \inst{1,2,3} \and Martin Lemoine\inst{1} \and
  Kumiko Kotera\inst{1,4} } 

\titlerunning{Rare transient sources of ultrahigh energy cosmic rays}

\authorrunning{S. Kalli, M. Lemoine \& K. Kotera}

\institute{ Institut d'Astrophysique de Paris,
  CNRS -- Universit\'e Pierre \& Marie Curie, 98 bis boulevard Arago,
  F-75014 Paris, France \and  Universit\'e Mohamed Boudiaf,
  Msila, Alg\'erie\and Laboratoire de Physique Mathematique et Physique Subatomique (LPMPS),  Universit\'e Mentouri, Constantine, Alg\'erie \and Department of Astronomy \& Astrophysics,  Enrico Fermi Institute, and Kavli Institute for Cosmological
  Physics, The
  University of Chicago, Chicago, Illinois 60637, USA.\\
  \email{kalli@iap.fr} } 

\abstract {Detecting and characterizing the anisotropy pattern of the
  arrival directions of the highest energy cosmic rays are crucial
  steps towards the identification of their sources. We
  discuss a possible distortion of the cosmic ray flux  induced by the anisotropic and inhomogeneous distribution of
  extragalactic magnetic fields in cases where sources of
  ultrahigh energy cosmic rays are rare transient phenomena, such as
  gamma-ray bursts and/or newly born magnetars.  This distortion does not  
  involve an angular deflection but the modulation of the flux related to the probability of seeing the source on an experiment 
  lifetime. To quantify
  this distortion, we construct sky maps of the arrival directions of
  these highest energy cosmic rays for various magnetic field
  configurations and appeal to statistical tests proposed in the
  literature. We conclude that this distortion cannot affect present
  experiments but should be considered when performing
  anisotropy studies with future large-scale experiments that record
  as many as hundreds of events above $6\times10^{19}\,$eV.}

   \maketitle
\section{Introduction}
The origin of ultrahigh energy cosmic rays (UHECR) -- particles with
energy $E\gtrsim 10^{19}\,$eV -- remains one of the longest running outstanding
problems of astrophysics. The extremely low flux that prevails at the
highest energies, $\sim 1/{\rm km}^2/{\rm century}$, makes it difficult
to accumulate more than a handful of events per year at $\sim
10^{20}\,$eV ($=\,100\,$EeV). The construction of large-scale
detectors has nevertheless brought a wealth of significant results
in the past few years. In particular, the Fly's Eye experiment
\citep{Abbasi2008,Abbasi2010} and the Pierre Auger Observatory
\citep{Abraham2008,Abraham2010} have detected a high energy cut-off at
$\sim 6\times 10^{19}\,$eV, in excellent agreement with the predicted
Greisen-Zatsepin-Kuzmin (GZK) cut-off \citep{Greisen,Zatsepin}, which
results from the pion production interaction of protons above that
energy with cosmic microwave background photons. Strictly speaking,
one cannot exclude at this stage that this cut-off represents the
maximal energy during acceleration, yet if one accepts the GZK
interpretation, this result would confirm that the sources of
ultrahigh energy cosmic rays are extragalactic and located within the
GZK sphere of radius $\sim 100\,$Mpc.  Similar arguments and
conclusions can be drawn for ultrahigh energy nuclei, although the
energy loss differs, being caused by the photodissociation of the nuclei.

The Pierre Auger Observatory has also announced the detection of
anisotropy at the 99\% c.l. in the arrival direction map of UHECR with
$E\gtrsim 6\times 10^{19}\,$eV \citep{Abraham2007,Abraham2008b}. This
anisotropy is expected if the (unknown) sources of ultrahigh
energy cosmic rays reside in large-scale structure -- which itself
appears anisotropic up to depths of several hundreds of Mpc --
provided that the magnetic deflection accumulated by the UHECR is not too
large. A detailed analysis of the anisotropy pattern indicates that it
is consistent with a distribution of UHECR sources across the local
large-scale structure \citep{Kashti,Koers,Aublin}.

If confirmed by future data -- this anisotropy is at present rejected
by the HiRes experiment in the northern hemisphere
\citep{2010ApJ...713L..64A} -- this result would be of prime
importance, notably because it would open the way to charged particle
astronomy. In particular, one could envisage probing the distribution
of the population of ultrahigh energy cosmic ray sources with future
large-scale cosmic ray detectors, a crucial step towards the
identification of the sources of these particles. This, of course,
would require a large effort on the experimental side, to
build detectors with an aperture significantly larger than already
achieved. On the theoretical side, it also requires the understanding of the
various biases that can distort our reconstruction of source
properties from sky maps of arrival directions.

Magnetic fields on large scales, Galactic and intergalactic, are prime
suspects in this regard, and many studies have been devoted to
characterizing their influence on the arrival directions (see the
discussion in \citealp{KL08b} and references therein). As our
understanding of cosmic magnetic fields has improved over the years,
the models have evolved. In particular, it is now understood that
the typical amount of deflection must vary according to the direction
in the sky, not only because the Galactic magnetic field exerts a
different influence in different directions (eg
\citealp{Alvarez-Muniz},\citealp{Takami}), but also because the
projected extragalactic magnetic field appears to be anisotropic, as
does the large-scale structure on the GZK sphere length scale.

In the present paper, we discuss a possible distortion of
the maps of arrival directions that is associated with the
inhomogeneous distribution of large-scale magnetic fields which
has so far not been discussed. If sources of ultrahigh energy cosmic
rays are rare and powerful events, such as gamma-ray bursts
\citep{Milgrom,Vietri,Waxman95} or newly born extragalactic magnetars
\citep{Arons}, the probability of detection of cosmic rays from a given source is extremely small in an experiment lifetime unless the arrival times of these cosmic rays have been substantially dispersed by intervening intergalactic magnetic fields. As we discuss in Section~2.1,
this implies that the flux in those directions of the sky with
small integrated column densities of magnetized gas should be low
relative to the mean background flux. To assess the
magnitude of this effect, we perform numerical Monte Carlo simulations
of the event detection, assuming that the source distribution of ultrahigh
cosmic rays follows the large-scale structure (traced using the PSCz
catalogue, \citealp{Saunders00}) and accounting for the inhomogeneous
distribution of intervening magnetic fields.

Our study indicates that the above-mentioned effect could only
marginally affect the results of current experiments such as the
Pierre Auger Observatory, but that it should be taken into account by
future experiments that might record as many as $10^2-10^3$ events
above $60\,$EeV. The magnitude of this effect is directly controlled by
the average depth to scattering against magnetized structures in the
intergalactic medium. The layout of the present paper is as
follows. In Section~2, we discuss the general conditions relating to
the source characteristics, the degree of magnetization of the
Universe for which the above effect applies and we specify how the
bias is modeled. In Section~3, we construct maps of arrival
directions of ultrahigh energy cosmic rays for various magnetic
configurations and we present quantitative estimates of the magnitude
of the effect for simulated sets of events. We draw our conclusions in
Section~4.

\section{Model}
\subsection{Magnetic field topology}
As discussed in detail in \cite{KL08b}, the distribution of magnetic
fields in the Universe, as experienced by a ultrahigh energy cosmic
ray, is quite likely to be highly inhomogeneous, being comprised of
magnetized structures such as filaments, halos of starburst galaxies,
lobes of radio-galaxies, and halos of clusters, separated by giant
unmagnetized voids. As long as the magnetic field
$B\lesssim10^{-12}\,$G in some region, this region can be considered
as unmagnetized from the point of view of the cosmic rays that
traverse it. We note that even extreme models that predict
magnetogenesis at high redshift with a strong magnetic field lead to
the above picture, of a Universe in which filaments are much more
strongly magnetized than the voids, as a result of the amplification of
the magnetic field during structure formation and the dilution of the
magnetic field in expanding voids.

In this situation, one may describe the propagation of the ultrahigh
energy cosmic rays as a sequence of stochastic interactions with
magnetized structures, which play the role of scattering
agents. Notwithstanding deflection by the Galactic magnetic field, the
overall effect is characterized in particular by the typical optical
depth for magnetic deflection, $\tau$. The optical depth is dominated
by the structures with the largest $n \sigma$, where $n$ represents
the space density and $\sigma$ the cross-section of the magnetized
halo.  Scanning through the possible scattering agents in the
Universe, \cite{KL08b} pointed out that the major protagonists in this
stochastic UHECR propagation model are: (1) halos of old radio-galaxies
with number density, typical radius and magnetic field in the range of
$n_{\rm rg}\sim 3\times10^{-3}-3\times10^{-2}$Mpc$^{-3}$,
$r_{\rm rg}\sim 1-3$~Mpc, and $B_{\rm rg}\sim 1-10\times 10^{-8}$G
respectively \citep{ME01,FL01}; (2) magnetized galactic winds with $n_{\rm
  gw}=2-5\times 10^{-5}$~Mpc$^{-3}$, $r_{\rm gw}=0.5-1$~Mpc, and
$B_{\rm rg}\sim 1-10\times 10^{-8}$G \citep{BVE06}; and/or (3) the
filaments/walls themselves if they have been magnetized by some
process or the filling factor of the above radio relics and
superwinds in the filaments reaches unity.  This can be summarized by
saying that magnetized scattering centers are distributed in
filaments/walls of large-scale structure -- with typical interaction
lengths of $d_{\rm f}\sim30\,$Mpc -- and that these magnetized structures
account for a covering fraction $\eta<1$ of the filament/wall
projected area.  The typical optical depth for magnetic scattering on
a GZK distance scale then reads
\begin{equation}
  \langle\tau\rangle\,\simeq\,3\eta\,
\left({l\over 100\,{\rm Mpc}}\right)\left({d_{\rm f}\over
      30\,{\rm Mpc}}\right)^{-1}\ .\label{eq:tau}
\end{equation}
It is important to recall that one might well find $\eta$ to be
significantly smaller than unity if the filling factor of magnetized
regions is small, as suggested for instance by the latest numerical
simulations of enrichment of the intergalactic medium reported
in~\cite{2009MNRAS.392.1008D}. Nevertheless, one can bound the average
optical depth below by the contribution of clusters of galaxies to
the scattering depth: with a density $n_{\rm c}\sim
10^{-5}\,$Mpc$^{-3}$, and a radius out to which magnetization can be
considered sufficient for UHECR deflection, $r_{\rm c}\sim\,$a few
Mpc, one finds that $\langle\tau\rangle_{\rm c} \,\sim\,0.08 (r/5\,{\rm
  Mpc})^2(l/100\,{\rm Mpc})$. We use a lower bound of $0.1$ for
$\langle\tau\rangle$ to a depth of $100\,$Mpc in what follows.

Since the source distance of protons of energy $\geq60\,$EeV
cannot exceed $200\,$Mpc -- comparable or smaller horizons are found for
heavy nuclei -- one expects the optical depth $\tau$ to that distance
to depend on direction angle, just as the projected large-scale
structure.

\subsection{UHECR bursting sources and inhomogeneous cosmic magnetic
  fields}\label{sec:general}
Henceforth, we assume that the sources of ultrahigh energy cosmic rays
are of the bursting type, with a rate per unit volume of $\dot n_{\rm
  s}$. If the sources were steady emitters, then the effect described
in this paper would not take place. The time delay imparted by
the magnetic field cannot modify the flux predictions of steady
emitters; the only effect of intervening magnetic fields is in this
case limited to deflection. Nevertheless, we note that there
are phenomenological arguments against the existence of steady sources of ultrahigh
energy cosmic rays (see \citealp{Lemoine09} for instance).

We consider a volume element $\delta V$ located at a distance $r$
from us, with an optical depth $\tau$ against magnetic
scattering. Cosmic rays emitted by sources in $\delta V$ have a
probability $p_\tau=1-\exp(-\tau)$ of suffering at least one
interaction with a magnetized system, hence of suffering a dispersion
$\Sigma_t$ in their arrival times, which can be calculated as follows.
One interaction with a structure of magnetic field strength $B$,
coherence length $\lambda_B$, and typical transverse size $r$, implies a
time delay of $\delta t\simeq 10^3\,{\rm yr}\,(r/2\,{\rm
  Mpc})^2(B/10\,{\rm nG})^2(\lambda_B/100\,{\rm kpc})(E/100\,{\rm
  EeV})^{-2}$ \citep{KL08b}; the total time delay then scales as the
number of interactions $N_{\rm int}$ (which is Poisson distributed
with a mean of $\tau$). As pion production occurs in a stochastic way,
the dispersion in the arrival times at energies above 60~EeV is comparable
to the total time delay (Waxman \& Miralda-Escude 1996, Lemoine et
al. 1997, Kotera \& Lemoine 2008), hence $\Sigma_t \sim N_{\rm
  int} \delta t$. The average dispersion for those particles that
experience at least one interaction can be straightforwardly evaluated as
$\Sigma_t = \tau\left[1-\exp(-\tau)\right]^{-1}\delta
t$. Consequently, a detector of area $A_{\rm exp}$ and exposure time
$T_{\rm exp}$ (with $T_{\rm exp}\ll\sigma_t$ expected) will record a
mean number of $N_{{\rm ev/s},\Sigma_t} = N_{\rm UHECR} A_{\rm exp}
T_{\rm exp} /(4\pi r^2\Sigma_t)$ from each source, assuming each
produces $N_{\rm UHECR}$ cosmic rays above some threshold energy
$E_{\rm thr}$. The mean number of sources contributing is $N_{\rm
  s,\Sigma_t}=\dot n_{\rm s}\delta V\Sigma_t$, so that the mean total
number of events received is $N_{\rm obs,\sigma_t} = N_{\rm
  UHECR}A_{\rm exp}T_{\rm exp} \dot n_{\rm s}\delta V/(4\pi r^2)$. As
expected, the dispersion in arrival times does not influence the
flux. As is well-known, the cosmic-ray energy output $E_{\rm UHECR}$
per source and the source density rate $\dot n_{\rm s}$ are related to
each other, in this case, through the normalization of the
experimentally determined cosmic ray flux: $E_{\rm UHECR}\dot n_{\rm
  s} \simeq 0.5\times 10^{-44}\,$erg/Mpc$^3$/yr (\citealp{Katz} for a
recent estimate).

In principle, cosmic rays emitted by these sources also have a
probability $1-p_\tau=\exp(-\tau)$ of traveling to Earth without
suffering an interaction with a magnetized system in the intergalactic
medium. However, from the point of view of an experiment, with
lifetime $T$, the detection of these particles is extremely unlikely if
the sources are rare and bursting, such as gamma-ray bursts.  For
instance, the occurrence rate of sources within a finite solid angle
$\Delta\Omega$, out to some distance $r$, is given by: $ \nu(\Delta\Omega) =
3\times 10^{-4}\,{\rm yr}^{-1}\, \Delta\Omega\,r_{100}^3\dot n_{\rm s,
  -9}$, where $\dot n_{\rm s, -9}=\dot n_{\rm s}/10^{-9}\,{\rm
  Mpc}^{-3}{\rm yr}^{-1}$, $r_{100}=r/100\,{\rm Mpc}$, and
$\Delta\Omega$ is expressed in steradians. For reference, the most
recent estimate of the local long gamma-ray burst rate is
$1.3\,+0.6\,-0.7\times 10^{-9}$Mpc$^{-3}$yr$^{-1}$
\citep{Wanderman}. The probability of detecting cosmic rays in that
area of the sky is then $T\nu(\Delta\Omega)\ll1$.

If such cosmic rays were detected, they would produce a giant flare in the detector, i.e., a localized anomalously large number of cosmic ray
events detected during a short period in time, 
 if $\dot n_{\rm s}$ is as low as that considered
above. In the absence of intergalactic deflection, the total
dispersion in arrival times indeed reduces to the contribution of the
turbulent component of the Galactic magnetic field; with a scale
height of $r_B\sim2\,$kpc \citep{Han}, strength $\delta B_{\rm MW}\sim
3\,\mu$G and coherence length $\lambda_B\sim 100\,$pc, one finds a
time delay $\delta t_{\rm MW} \simeq 0.2\,{\rm yr} (r_B/2\,{\rm
  kpc})^2(\delta B_{\rm MW}/3\,\mu{\rm G})^2(\lambda_B/100\,{\rm
  pc})E_{60}^{-2}$ and $\sigma_{t,\rm MW}\lesssim \delta t_{\rm MW}$,
assuming that these cosmic rays are protons (see further below for heavier
nuclei). Cosmic rays can be expected to have a similar dispersion in
arrival times upon exiting the host galaxy of their source. Assuming
that the total time dispersion for these particles is $\delta t_{\rm
  MW}\,\ll\,T$, the mean number of events per source is
$N_{{\rm ev/s},0} = N_{\rm UHECR} A_{\rm exp} /(4\pi r^2)$, much
larger than expected according to the previous limit by a factor
of $\Sigma_t/T_{\rm exp}$. As to the mean number of sources contributing
to the flux, one finds $N_{\rm s,0}=\dot n_{\rm s}\delta V T_{\rm
  exp}$. To provide quantitative estimates, we write $E_{\rm
  UHECR}=10^{53}\,E_{53}\,{\rm erg}$ ($E_{53}\sim1$ provides the
correct normalization of the flux for $\dot n_{\rm
  s}\sim10^{-9}\,$Mpc$^{-3}$yr$^{-1}$) and estimate that $N_{{\rm ev/s},0}
\sim 3\times 10^4\, E_{53}r_{100}^{-2}$ (for $A_{\rm 
  exp}=3000\,$km$^2$ corresponding to the Pierre Auger Observatory).
No such flare has been observed by the Pierre Auger Observatory, for
which the typical number of events per source above 60~EeV is on the order of
unity or less \citep{Abraham2008}.

When such flares are not detected, and one assumes that
the sources are rare and bursting, then the arrival direction flux
must be modulated by the probability of suffering at least one
interaction with a magnetized system, which is $p_\tau=1-\exp(-\tau)$
as discussed above. We note that this modulation might be even more
extreme if the particle had to encounter more than one scattering
event in order to disperse its arrival time sufficiently for it to be
detectable; we adopt the above (slightly conservative)
modulation in the following.

The above modulation formally corresponds to the conditional
probability of detection given that no flare has been
detected (under our model assumption that the source is bursting, of
rare occurrence). It can also be seen as the consequence of a hierarchy
of timescales. At a continuous rate, the detector registers particles
that have experienced at least one interaction with magnetized systems in
the intergalactic medium, provided that the occurrence rate $\nu$ of
bursting sources in the GZK sphere and the average dispersion across that
distance scale satisfy $\Sigma_t\nu\,\gg\,1$. However, the arrival
direction map of these particles is modulated by $p_\tau$. To
detect flares of particles that have traveled without interacting
with magnetized structures, one would need to integrate over a
timescale $\nu^{-1}$, but $\nu^{-1}\,\gg\,T_{\rm exp}$ for rare
bursting sources.

Obviously, if the energy output of each source is small enough (see
below), i.e. if the source occurence rate is sufficiently high, the flares
detected without intergalactic deflection could
produce fewer than one event in the detector, in which case one could
not tell whether intergalactic deflection has taken place, and the
modulation $1-\exp(-\tau)$ should not be applied. The
latter statement depends of course directly on the exposure of the
instrument. For the Pierre Auger Observatory, for instance, if no significant clustering is found \citep{Abraham2008}
then no such flare (i.e. without intergalactic time
dispersion) has been detected from bursting sources with energy output
$E_{\rm UHECR}\gtrsim 10^{49}\,$erg. This implies that for sources
with an occurrence rate $\dot n_{\rm s}\,\ll\,
10^{-5}\,$Mpc$^{-3}$yr$^{-1}$, the above modulation $1-\exp(-\tau)$
should be taken into account when comparing the arrival directions
of ultrahigh energy cosmic rays with possible source distributions.  For
future instruments with an aperture ten times that of the Pierre Auger
Observatory, this upper bound on the density rate should be increased
tenfold.

To summarize, the distortion of arrival directions maps that we
examine in the following Section takes place for bursting sources that
rarely occur in the GZK sphere, as measured comparatively to an
experiment lifetime. Gamma-ray bursts sources of ultrahigh energy
cosmic rays and magnetars fall into this category. Regarding the
latter, the typical energy output is $\sim 10^{51}\,$ergs at
$10^{20}\,$eV for a typical rate of $10^{-5}$/yr per galaxy,
i.e. $\sim 10^{-7}\,$Mpc$^{-3}$yr$^{-1}$ \citep{Arons}. The
acceleration of ultrahigh energy cosmic rays in blazar flares has also
been discussed \citep{Dermer}; however, in this scenario the
particles experience a substantial time delay $\sim10^5\,$yr upon exiting the
lobes of the associated radio-galaxy, hence the source appears as a
steady emitter of ultrahigh cosmic ray sources for all phenomenological
purposes. The effect that we discuss therefore does not apply to these
models.

One cannot exclude that ongoing or future detectors could detect a
giant flare associated with the observation of events that have not
suffered dispersion in their arrival times, all the more so at source
occurrence rates of $\dot n_{\rm s}\sim 10^{-6}\,$Mpc$^{-3}$yr$^{-1}$ and
small depth to magnetic scattering of $\tau<1$. The typical rise and
decay timescale then corresponds to $\lesssim \delta t_{\rm MW}$, a
fraction of a year. In this case, one would be able to asses
directly the cosmic ray energy output of the source, along with the
possible detection of a counterpart, which would provide unvaluable
information.

Finally, we should mention that we have assumed that ultrahigh energy
cosmic rays are protons.  There is currently no clear determination of
the chemical composition of ultrahigh energy cosmic rays. While the
HiRes experiment has reported a composition with a proton fraction
that increases beyond the ankle \citep{Abbasi2010}, the measurements
of the Pierre Auger Observatory point instead toward a composition that
becomes increasingly heavier beyond the ankle \citep{Abraham2010}. By
themselves, these results of the Pierre Auger Observatory are somewhat
puzzling because they may be in apparent contradiction with the anisotropy
results (see e.g. \citealp{Lemoine09}) and no hadronic model
of shower reconstruction has so far been able to reproduce all of the
composition data of the Pierre Auger Observatory \citep{Abraham2010}.
Nevertheless, if the composition were predominantly heavy with charge
$Z\,\gg\,1$, then the above discussion would remain valid, but the sky
maps discussed below would not be realistic owing to the large
deflection imparted by the Galactic magnetic field. In practice, our
imperfect knowledge of the Galactic magnetic field would limit the
impact of the \textbf{distortion} discussed below.

\subsection{Flux calculation}\label{sec:fluxcalc}
To model the flux and simulate the events, we adapt the
method introduced by \cite{Waxman97}: in a given volume element
$\delta V$, the mean number of cosmic ray sources is written $N_{{\rm
    s},\Sigma_t}=\dot n_{\rm s}\Sigma_t \delta V$, where
$\Sigma_t=\tau \delta t/\left[1-\exp(-\tau)\right]$ on average,
following section~\ref{sec:general}; in our simulations, we neglect
the random nature of $\Sigma_t$, which can be regarded as a higher
order effect as its magnitude does not modify the average flux, only
the random fluctuations around this average. The mean number of cosmic
ray events emitted by each source is then, following the earlier
discussion, $N_{{\rm ev/s},\Sigma_t} = N_{\rm UHECR} A_{\rm exp}
T_{\rm exp} /(4\pi r^2\Sigma_t)$ and the mean contribution to the flux
from this region is therefore $N_{{\rm ev/s},\Sigma_t}N_{{\rm
    s},\Sigma_t}$. To account for multiple events per source,
one considers the probability that $S_i$ sources produce $i$ events;
these $S_i$ variables are indeed independent of each other and
Poisson distributed with mean $N_{{\rm s},\Sigma_t}P_i$, where
$P_i=N_{{\rm ev/s},\Sigma_t}^i\exp\left(-N_{{\rm
      ev/s},\Sigma_t}\right)/i!$ is the Poisson probability that a
given source produces $i$ events \citep{Waxman97}.

One can use the above direct averages multiplied by $1-\exp(-\tau)$ to compute an average flux distribution, as we do in
Section~2.3.1. Direction-dependent effects are incorporated as
follows. We use the following law to express the optical depth at
distance $r$ in direction $\mathbf{n}$
\begin{equation}
  \tau(r,\mathbf{n})\,=\,\langle\tau\rangle\int_{0}^{r} {\rm d}l \,
    \frac{\rho_{\rm g}(l,\mathbf{n})}{\langle\rho_{\rm g}\rangle}\ ,
\end{equation}
which amounts to scaling the density of magnetized structures
according to the galaxy density $\rho_{\rm g}(l,\mathbf{n})$ (at
distance $l$, in the direction $\mathbf{n}$). The average optical
depth against scattering with magnetized structures
$\langle\tau\rangle$, is calculated by averaging over all sky, and
bears the typical values calculated in Eq.~(\ref{eq:tau}). In what
follows, we use the definition $\langle\tau\rangle_{100}$ to express
this quantity calculated to a depth of $100\,$Mpc. The galaxy density
is traced through the PSCz survey of galaxies \citep{Saunders00}. We
also assume that the density of sources follow the galaxy density as
traced by the PSCz survey. These choices are motivated by the PSCz being a good tracer of star-forming galaxies, which should
host bursting sources of ultrahigh energy cosmic rays such as
gamma-ray bursts and magnetars, and contribute to the
magnetic enrichment of the intergalactic medium. We then integrate the
contributions of each volume element along $r$ in each direction of
the sky, modulated by the appropriate $1-\exp(-\tau)$ probability of
intersection with magnetized structures, to compute the map
of arrival directions.

We note that, for a given average source density $\dot n_{\rm
  s}$, the final average number of events detected depends on the
average optical depth as a consequence of the flux modulation 
by $1-\exp(-\tau)$ in each direction. To produce comparable sets
of events, we thus rescale the density $\dot n_{\rm s}$ as a function
of $\tau$ in order to keep the average number of events fixed as
$\tau$ varies. It is straightforward to see that the all-sky average
flux out to a distance $d$ scales with average magnetic depth $\tau$ as
$\left[1-1/\langle\tau\rangle+\exp(-\langle\tau\rangle)/\langle\tau\rangle\right]$
(with $\langle\tau\rangle$ calculated up to depth $d$) if one assumes
a homogeneous and isotropic distribution of sources and magnetized
structures; if $\langle\tau\rangle\ll1$, the flux is thus suppressed
by $\langle\tau\rangle/2$, but the suppression factor converges to
unity as $1- 1/\langle\tau\rangle$ for $\langle\tau\rangle>
1$. Therefore, for an average optical depth of order unity on the
horizon distance scale for particles above the threshold energy
$E_{\rm thr}$, the source density must be renormalized by a factor of
order unity. This renormalization lies well within the margins of the uncertainties
in the absolute rate of possible sources.

To simulate a finite sets of events, we resort to Monte Carlo
simulations, in which we rely on the above probability laws for $S_i$:
in each volume element of the sky, we draw the random variable $S_i$,
which characterizes the number of sources contributing $i$ events. The
total number of events detected from that volume element is then
$\sum_{i=1}^{\inf} i S_i$. In order to account for the modulation
effect by $p_\tau$, we assume that the mean number of sources
contributing to the flux in that volume element is $p_\tau N_{{\rm
    s},\Sigma_t}$.

\subsubsection{Mean flux} \label{sec:meanflux} 

The cosmic ray (CR) differential flux produced by sources at distances
smaller than $r_{\rm max}$ in direction $\mathbf{n}$ is given by
\begin{equation} 
  \mathcal{F}(\mathbf{n}) = 
\,\frac{\dot{N}}{4\pi} \int_{0} ^{ r_{\rm max}} {\rm d} l\ n_{\rm s}\ b(l,\mathbf{n})  \,,
\label{eq:flux}
\end{equation}
where $\dot{N}$ is the injection rate of particles by each source, $b$
denotes the bias of the source distribution, which we relate -- as
explained above -- to the galaxy density, i.e. $b (l,\mathbf{n}) =
\rho_{\rm g}(l,\mathbf{n})/\langle \rho \rangle$, $\langle \rho
\rangle$ is the mean density of the PSCz catalog, $n_{\rm
  s}=\Sigma_t\dot n_{\rm s}$ is the average apparent source density,
and $\Sigma_t$ is the typical dispersion in arrival times (which we keep its average value
fixed to, as discussed above). We consider two values of the apparent source density:
$n_{\rm s} = 10^{-3}\rm Mpc^{-3}$ and $n_{\rm s} = 10^{-5}\rm
Mpc^{-5}$. The former corresponds to the number density of bright
galaxies and the second is a lower bound inferred from experimental
data, as follows. First of all, the distance traveled by the highest
energy event ($E > 10^{20} $eV) detected by the Fly's Eye experiment
\citep{Bird} cannot exceed 50~Mpc; its detection thus indicates that
at least one CR source exists in the field of view of Fly's Eye out to
50 Mpc, implying a source density $n_{\rm s} \gtrsim 10^{-5}\rm
Mpc^{-3}$. Furthermore, the non-detection of significant clustering by
the Pierre Auger Observatory \citep{Abraham2008b} also implies that $n_{\rm
  s}\,\gtrsim\,10^{-5}\,$Mpc$^{-3}$ (see the discussion in
\citealp{Kashti}). 

These two different apparent source densities lead to different
predictions of the detection of multiplets of events. For $n_{\rm
  s}=10^{-5}\,$Mpc$^{-3}$ for instance, there are fewer sources in the
GZK sphere, hence for a projected sample of $N_{\rm ev}=100$ particles
of energy above 80~EeV from sources located within a distance below
$r_{\rm max}=100$Mpc, the number of apparent sources is $\sim 40$
leading to the appearance of multiplets, which can be used to
constrain the apparent source density. If $n_{\rm s} = 10^{-3}\rm
Mpc^{-3}$ however, the number of sources is $\sim4\times 10 ^3$, hence
the average number of event per source is smaller than unity. For an
energy threshold of 60~EeV, the horizon increases to $200$\,Mpc, hence
for a sample of 1000 particles as considered further below, the
conclusions are similar.  

We expect to detect more data from future experiments, such as Auger
North and JEM-EUSO. The development of Auger North is due to start in
2011 in Colorado (USA); this experiment should reach an area of more
than $20000\,$km$^2$, which represents an increase of about nearly an
order of magnitude when compared to Auger South \citep{AugerNorth}.
The Extreme Universe Space Observatory onboard the Japanese Experiment
Module (JEM-EUSO) is to be developed on the International Space
Station in 2014 \citep{JEMEUSO}.  With a field of view of
$60^{\circ}$ from a 430~km orbiting altitude, the integrated exposure time should
exceed $10^6\,$km$^2\,$sr$\,$yr. One should thus expect to detect more
than $10^3$ particles above $7 \times 10^{19}$ eV during its five year
operation.  For this experiment, the non-observation of multiplets of
events of energy above 80 EeV would indicate that $n_{\rm s} \gtrsim
10^{-3}$Mpc$^{-3}$.

In our simulations, the flux is normalized to the expected total flux
above a threshold energy $E$, i.e. $ N_{\rm tot}(> E) = \,\int A_{\rm
  exp}T_{\rm exp} f(\Omega) {\mathcal{F}(\mathbf{n})} {\rm d}\Omega $,
where as before, $A_{\rm exp}$ denotes the effective detector area,
$T_{\rm exp}$ the observation time, and $f(\Omega)$ the aperture of
our fiducial experiment.

\section{Simulations}

\subsection{Sky maps}

\begin{figure}[th]
\centering

\includegraphics[angle = 90,width=9cm]{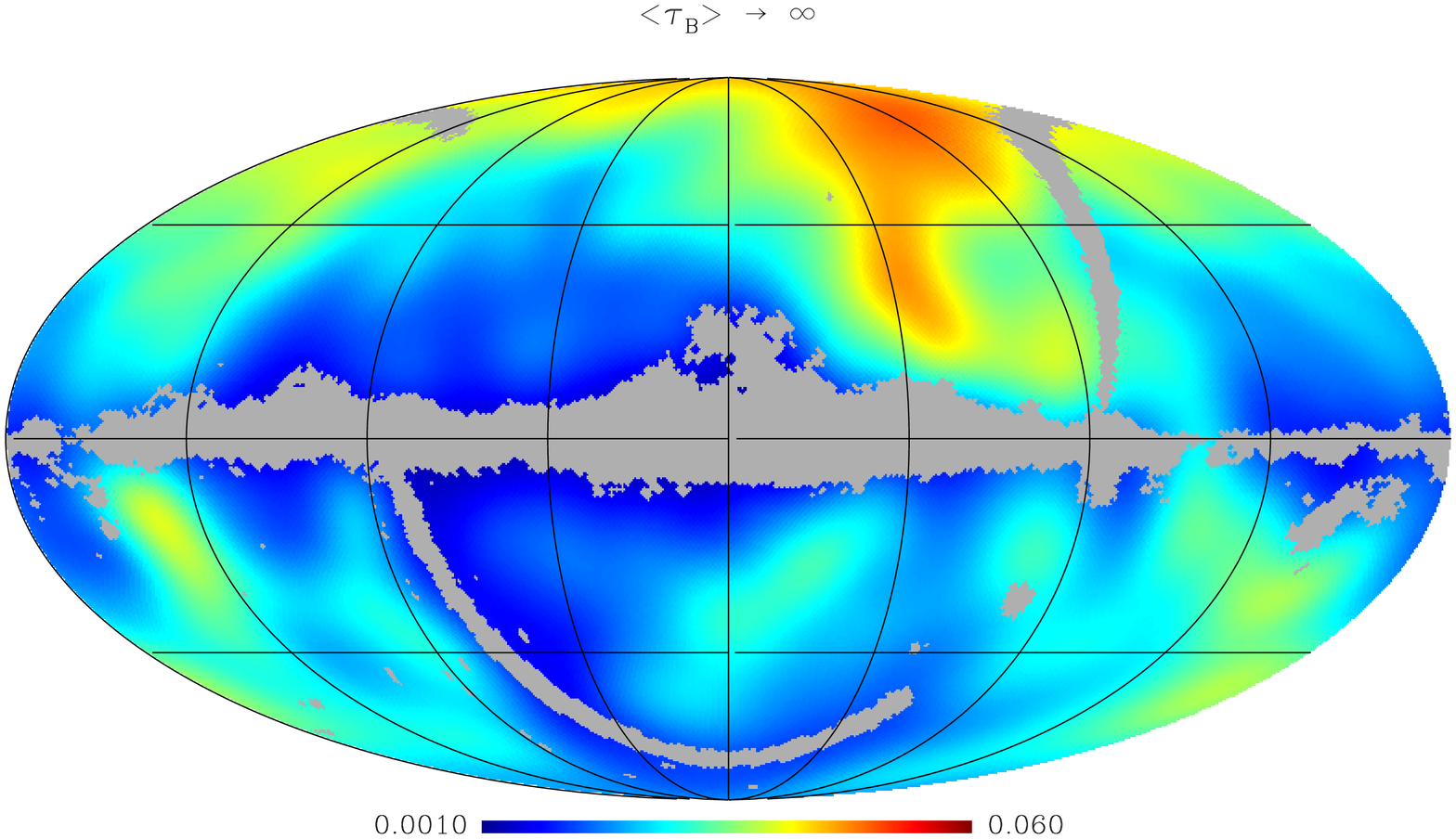}
\includegraphics[angle = 90,width=9cm]{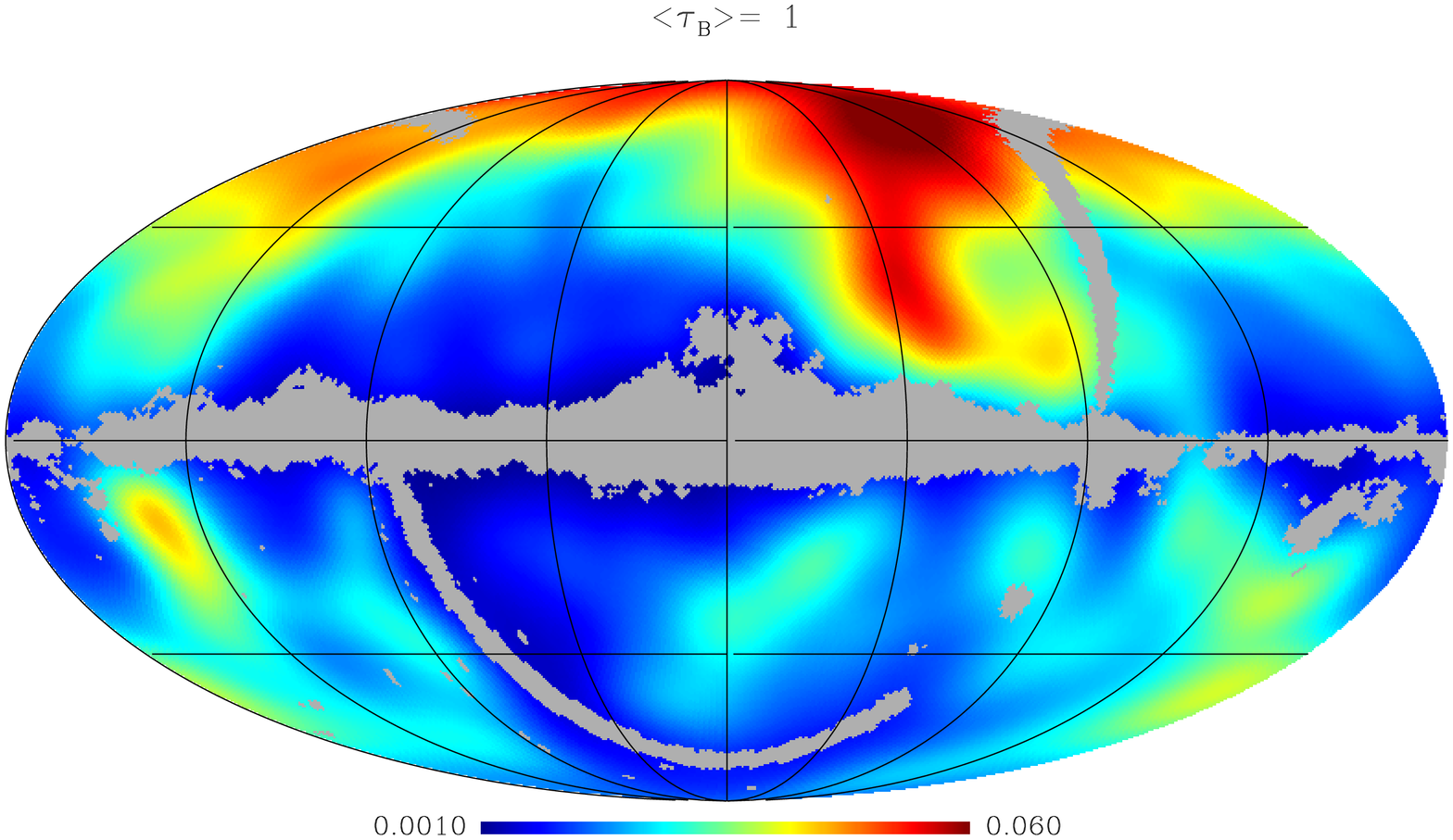}
\caption{ Sky maps in Galactic coordinates of the UHECR differential
  flux indicated in Eq.~(\ref{eq:flux}) for an energy threshold $E$ =
  60 EeV corresponding to a depth of 200\,Mpc. Top panel: LSS model,
  in which sources follow the matter distribution according to the
  PSCz survey and in which $\langle\tau\rangle_{100}\rightarrow
  +\infty$, corresponding to the assumption that the probability of
  scattering on a magnetized system is unity; bottom panel: same,
  but with $\langle\tau\rangle_{100} = 1$. The grey zone represents
  the mask of the PSCz catalog in which there is no information.  }
\label{fig:flux200}
\end{figure}
\begin{figure}[th]
\centering

\includegraphics[angle = 90,width=9cm]{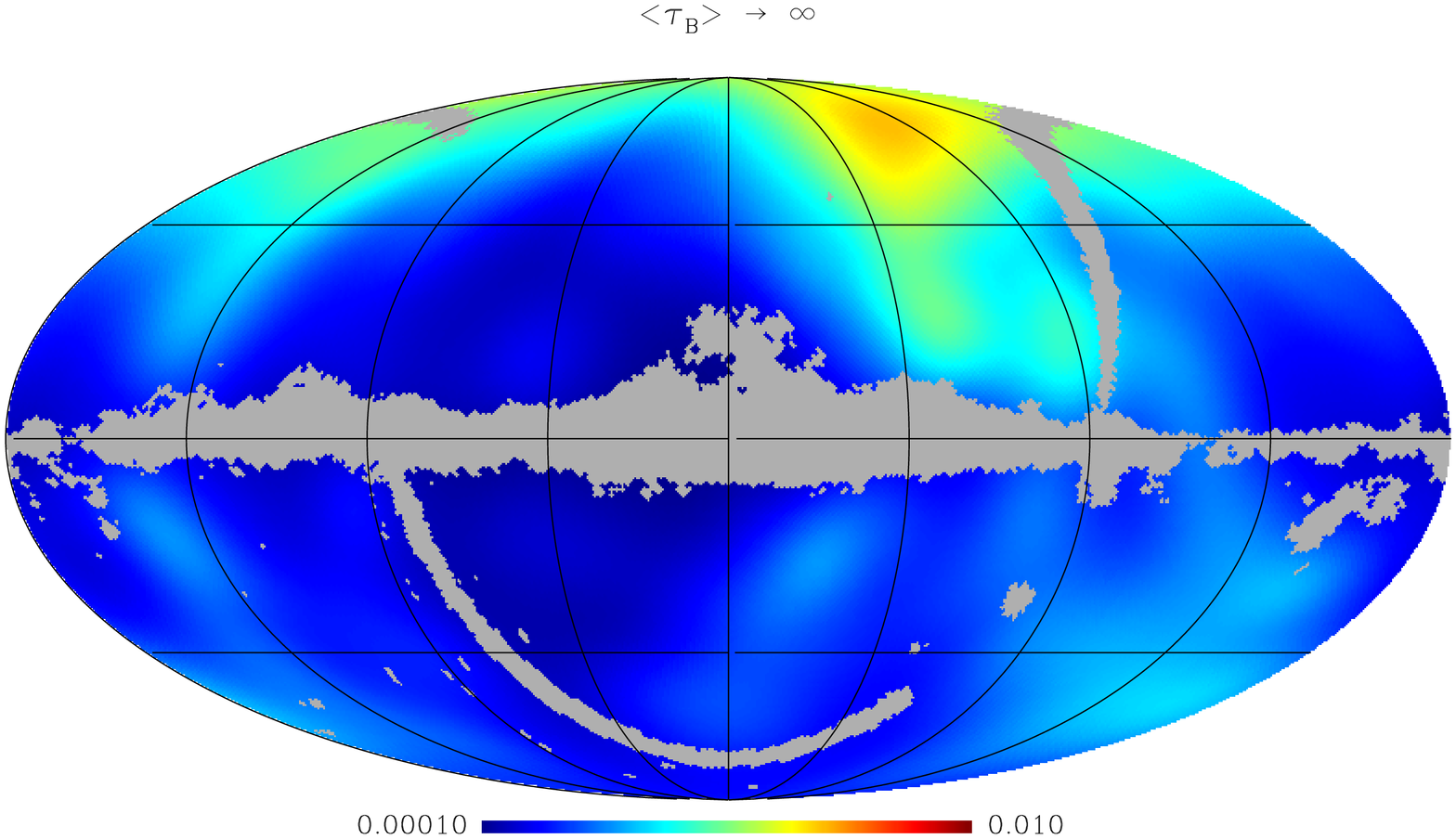}
\includegraphics[angle = 90,width=9cm]{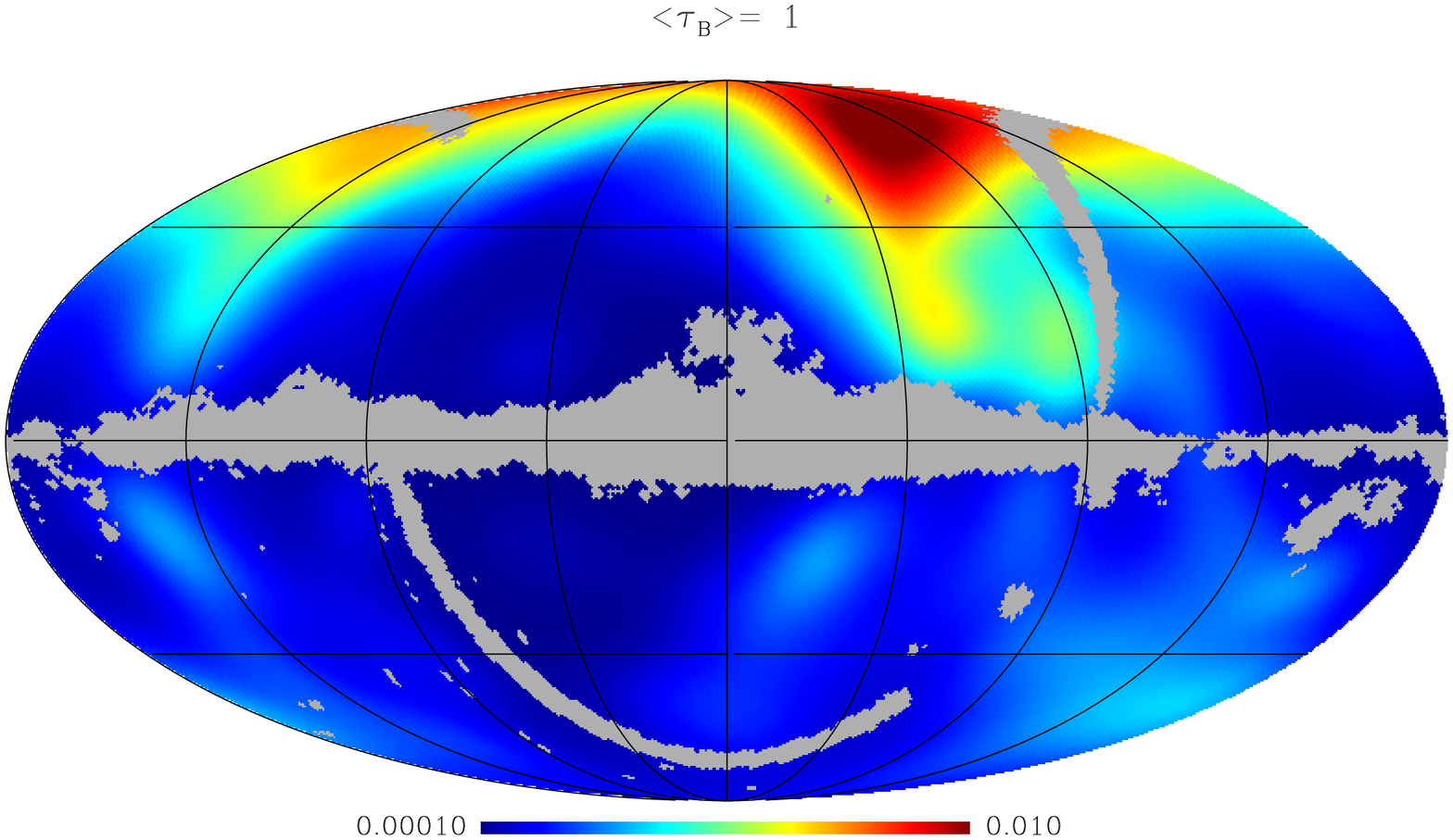}
\caption{ Same as Fig.~\ref{fig:flux200}, for an energy threshold
  $E$ = 80 EeV, corresponding to a depth of $100\,$Mpc. }
\label{fig:flux100}
\end{figure}

Using Eq.~(\ref{eq:flux}), we construct sky maps of the average UHECR flux
for two energy thresholds: 60~EeV and 80~EeV, corresponding  to integration depths of $200\,$Mpc and $100\,$Mpc,
respectively.
These sky maps, that do not include any experiment exposure factor for
completeness, are shown in Figs.~\ref{fig:flux200} and
\ref{fig:flux100} respectively. In each figure, the top panel
shows the mean flux expected, if the sources follow the large-scale
structure distribution and the modulation $p_\tau$ discussed above
is ignored. In the bottom panel, a modulation of
$\langle\tau\rangle_{100}=1$ has been adopted. Plotting this mean
flux corresponds formally to what would be observed for a continuous
distribution of sources (i.e. infinite apparent source density). The
top panel, with no modulation, indicates what would be seen if the
probability of scattering on the magnetized regions were everywhere unity,
i.e. $\langle\tau\rangle_{100}\rightarrow +\infty$, or if the sources
were steady emitters. The bottom panel shows the same map, but for bursting sources and including the
modulation by the patchy distribution of magnetic field in the local
Universe. For the sake of clarity, we have
constructed these plots with a relatively narrow gaussian filter in
HEALPix \citep{healpix} in order to smooth the galaxy distribution to
a relatively high resolution $\sim 1^{ \circ}$. This resolution is
higher than that of the PSCz catalog (about $7^\circ$) and smaller
than the typical deflection expected for the interaction with
magnetic fields. In the actual simulations discussed below, we used a
degraded filter with $4^\circ\times4^\circ$ pixels.

To take into account the effect of the magnetic configuration
on the received flux, we multiply the mean flux from each individual
cell (in three-dimensional Galactic coordinates $r,l,b$) by the
probability $p_{\tau}$. We use several values of $ \langle \tau
\rangle _{100}$: $0.1, 1, 3$, and $\infty$. We then obtain different
models of CR distribution depending on the optical depth value. In the
case $ \langle \tau \rangle \rightarrow \infty$, the probability
$p_{\tau}$ is equal to 1 and we are able to observe all p particles roduced
from powerful bursting sources that follow the large-scale
structure (LSS). Hereafter, we refer to this configuration as ``the
LSS model''. The two other magnetic configurations will be named
according to the value of the optical depth, e.g.  $\langle\tau\rangle
= 3$ model and $\langle\tau\rangle = 1$ model.

Figures \ref{fig:flux200} and \ref{fig:flux100} present the sky
maps of the UHECR differential flux normalized over the total expected
flux for UHECR energy thresholds $E = 60$~EeV and $E = 80$~EeV. Generally
speaking, the distribution of the flux reflects the galaxy
distribution following the LSS, but in the bottom map (corresponding
to $ \langle \tau \rangle =1$) the flux seems to be lower than in the top panel (corresponding to the LSS model, $ \langle
\tau \rangle _{100} \rightarrow \infty$); in the bottom panel, the
smaller number of scattering centers in the voids concentrate the flux
around regions of overdensity such as the Virgo cluster ($b\sim
74^{\circ},l\sim -80^{\circ}$), the Shapley cluster ($b\sim
29^{\circ},l\sim -54^{\circ}$), the Centaurus cluster ($b\sim
21^{\circ},l\sim -58^{\circ}$), the Hydra cluster ($b\sim
26^{\circ},l\sim -101^{\circ}$), the Pavo-Indus cluster ($b\sim
-23^{\circ},l\sim -28^{\circ}$), the Fornax cluster ($b\sim
-53^{\circ},l\sim -124^{\circ}$), and the Perseus Pices cluster ($b\sim
-17^{\circ},l\sim 124^{\circ}$), with a concomittant reduction in the
underdense areas of the sky. The general effect of the distortion that
we discuss here is thus to cluster the arrival
directions more strongly in the densest areas of the sky.

\subsection{Statistical tests}
To quantify the observed effect on the sky maps, we use
various statistical tests that have been discussed in the literature and are
 related to the anisotropy of ultrahigh energy cosmic ray arrival
directions: the $\chi^2 $ test \citep{Cuoco07}, the one-dimensional
Kolmogorov-Smirnov test \citep{Koers}, the two-dimensional
Kolmogorov-Smirnov test \citep{Harari}, the $Y$ test \citep{Koers},
the correlation test $X_{C}$ \citep{Kashti}, and the Kuiper
test. Although we tested them all, we present in what follows the
results of those tests that have proven to be the most sensitive: the $X_{C}$
test, the two-dimensional Kolmogorov-Smirnov test, and the Kuiper test. The $X_{C}$ test was found in \cite{Kashti} to be more sensitive to the anisotropy signature than the power spectrum and the two-point correlation function. The two-dimensional Kolmogorov-Smirnov test was used in \cite{Harari} to exclude isotropy when studying the UHECR distribution. And finally the Kuiper test, which is similar to the Y test defined in \cite{Koers}, was found to have the highest statistical power of all considered statistical tests when considering a small number of high energy events. We tried to span different types of tests: the $X_{C}$ test relies on binning the sky, and the Kuiper and two-dimensional Kolmogorov-Smirnov tests depend on cumulative functions of the flux, the first being one dimensional,  the second involving the Galactic coordinates \textit{ l} and \textit{ b}. Using these tests, we compare the distributions of simulated events from different models to the average distributions of events in the LSS model.
 (i.e. $ \langle \tau \rangle _{100}
\rightarrow \infty$).

As explained in Section~\ref{sec:fluxcalc}, we generate two sets of
simulated events for each model, corresponding to two possible
threshold energies $E=60\,$EeV and $E=80\,$EeV, i.e. a set of $N_{\rm ev}$ = 100
particles sampled neglecting the contribution from sources beyond $100$ Mpc (corresponding to the horizon for particles above $80\,$EeV)
and a set of $N_{\rm ev}$ = $1000$ particles from possible sources within $200$ Mpc
(corresponding to the horizon for particles above $60\,$EeV).

For the $X_{C}$ test, we divide the sky into angular bins of size $4
^\circ \times 4 ^\circ$ to avoid the local magnetic field effect as
mentioned above and use two reference models: the isotropic model
in which the arrival directions are purely isotropic (in the absence
of experimental aperture sensitivity) and the LSS model with $\langle
\tau \rangle _{100} \rightarrow \infty$. We then compare the number of
simulated events per bin in the tested model to average numbers of
events from the two reference models following the formula
\begin{equation} \label {eq:Xc} 
X_{C} =\,\sum_{i=1}^{N_{tot}}
  \frac{(N_i ^\tau -\langle N_{i,\rm LSS}\rangle)(\langle N_{i,\rm
      iso}\rangle -\langle N_{i,\rm LSS}\rangle)}{\langle N_{i,\rm
      LSS}\rangle}\,,
\end{equation} 
where $N_i ^\tau$ denotes the number of simulated events in angular
bin $i$ from the model to be tested with magnetic optical depth
$\tau$, $\langle N_{i,\rm LSS}\rangle$ denotes the average number of
events expected in angular bin $i$ in the reference LSS model and
$\langle N_{i,\rm iso}\rangle$ corresponds similarly to the average
number of events expected in angular bin $i$ in the isotropic model.

For the other two tests, we take as a reference model the LSS model and
use cumulative distributions of the flux.  The Kuiper test measures the quantity
$V$, which is defined as the maximum deviation above and below the two
considered cumulative distributions. Following \cite{Koers}, we name
$ \mathcal{C}_{\rm rand}(\mathcal{F})$ the cumulative distribution of
the simulated integrated flux from the model to be tested and
$\mathcal{C}_{\rm total}(\mathcal{F})$ the cumulative distribution of
the expected integrated flux from the reference model. We then have
\begin{eqnarray} \label {eq:kuiper}
V = \,\max[\mathcal{C}_{\rm rand}(\mathcal{F}) - \mathcal{C}_{\rm total}(\mathcal{F})]  
              +\max[\mathcal{C}_{\rm total}(\mathcal{F}) - \mathcal{C}_{\rm rand}(\mathcal{F})]  \,.
\end{eqnarray} 

For the two-dimensional Kolmogorov-Smirnov test, each simulated event
$n$ is characterized by its galactic coordinates ($l_n$,\,$b_n$); the
sky plane is then divided into four quadrants close to this direction
$\left(l< l_n,b<b_n\right)$, $\left(l> l_n,b<b_n\right)$, $\left(l<
  l_n,b>b_n\right)$, and $\left(l> l_n,b>b_n\right)$ and the test computes
the differences between the cumulative distributions of the tested and
reference models in each of the four quadrants. This procedure is
repeated $ 4 \times N_{ev}$ times and $D_{ks}$ is defined as the
maximum of these differences (between the cumulative distributions of
the tested and reference models). In other words,
\begin{equation}\label {eq:ks2d}
D_{KS} = \,\max_{Q=1..4} [|\mathcal{C}_{\rm rand}^{Q}(l,b) - \mathcal{C}_{\rm total}^{Q}(l,b)   |]\,,
\end{equation}
where $(l,b)$ are the Galactic coordinates of the cosmic ray positions
and $Q$ is one of the four quadrants with respect to a simulated event
position in which data can be accumulated.\\

\begin{figure}[th]
\centering

\includegraphics[angle = 90,width=9cm]{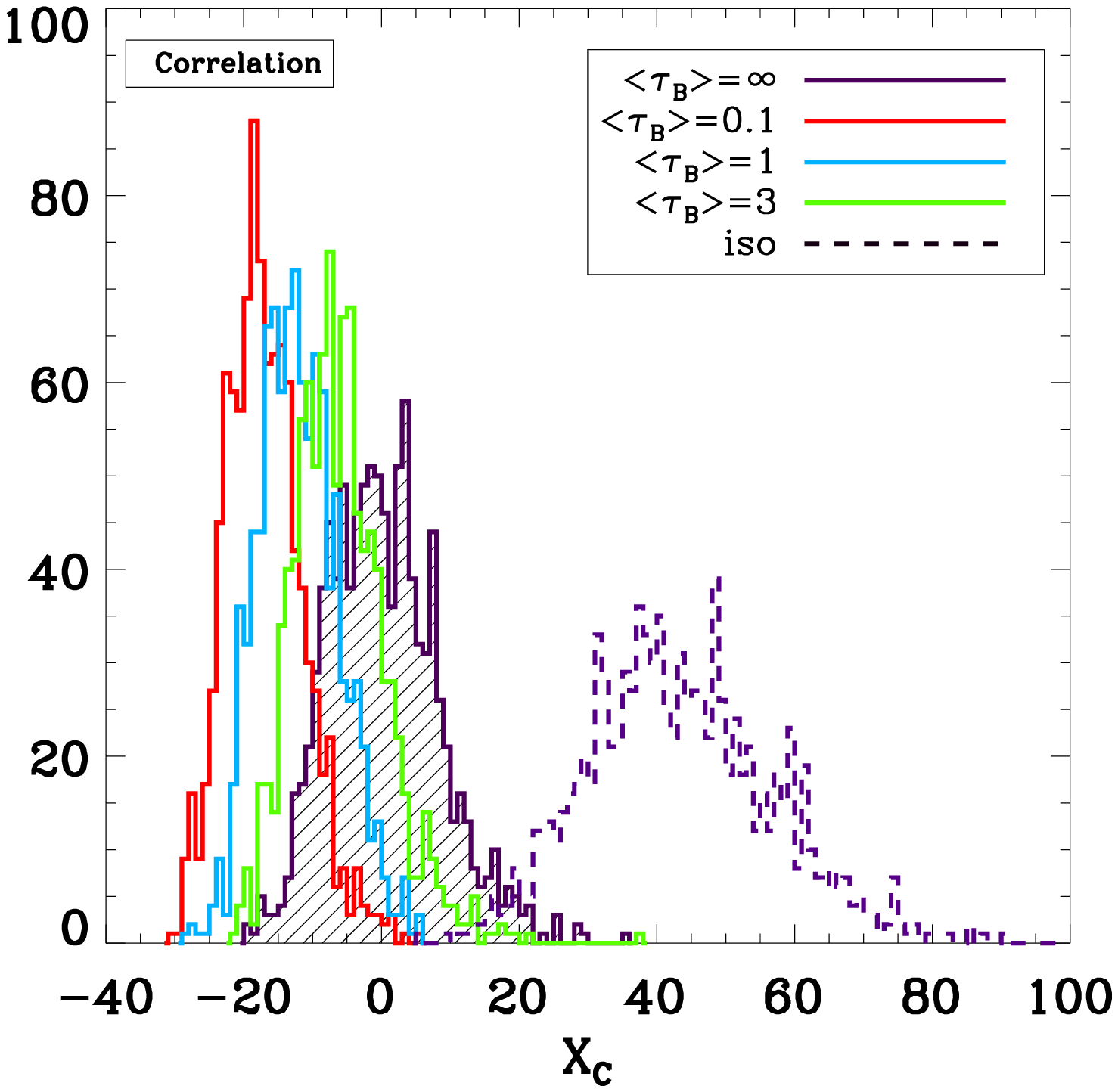}
\includegraphics[angle = 90,width=9cm]{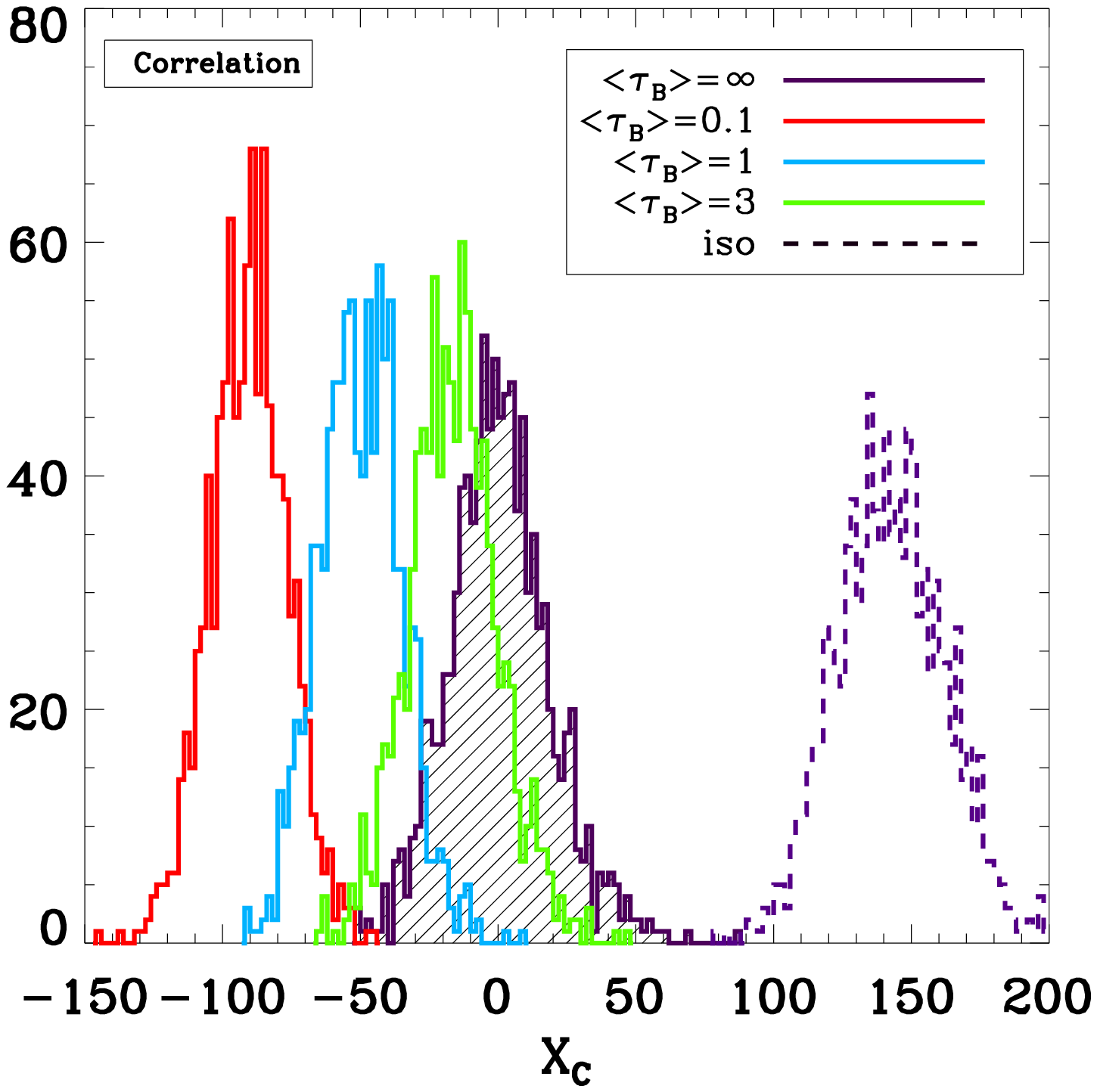}
\caption{Distribution of $X_{C}$ from 1000 Monte Carlo realisations of
  100 simulated events with energy above 80 EeV (top) and 1000
  simulated events with energy above 60 EeV (bottom). Mean source
  density $n_{\rm s} = 10^{-3}\rm Mpc^{-3}$. The dotted curves represent the
  isotropic model, the hatched curves denote the LSS model and from
  left to right we have $\langle\tau\rangle_{100}= 0.1$,
  $\langle\tau\rangle_{100}= 1$, and $\langle\tau\rangle_{100}= 3$.}

\label{fig:histo_cor}
\end{figure}

\begin{figure}[th]
\centering

\includegraphics[angle = 90,width=9cm]{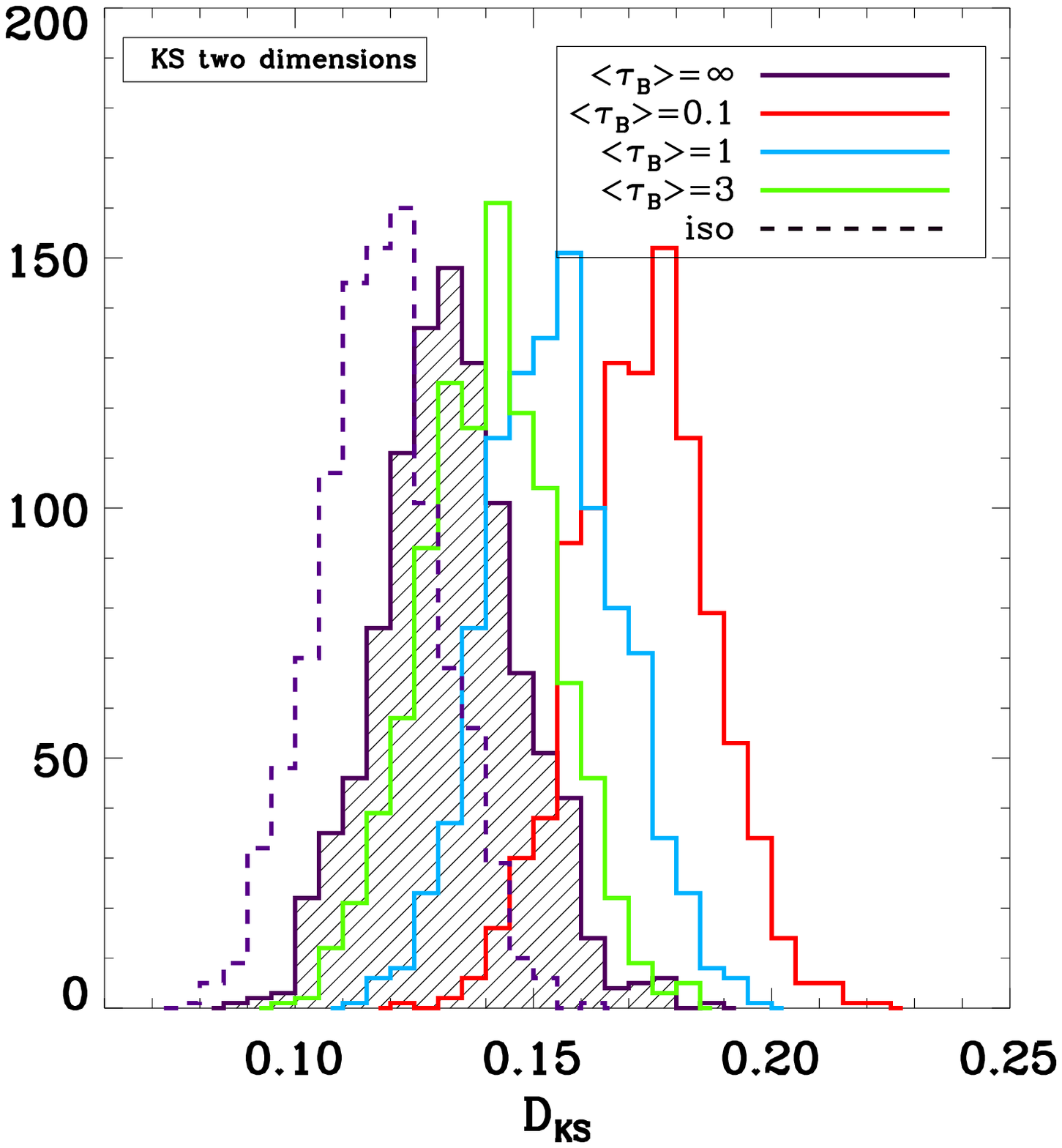}
\includegraphics[angle = 90,width=9cm]{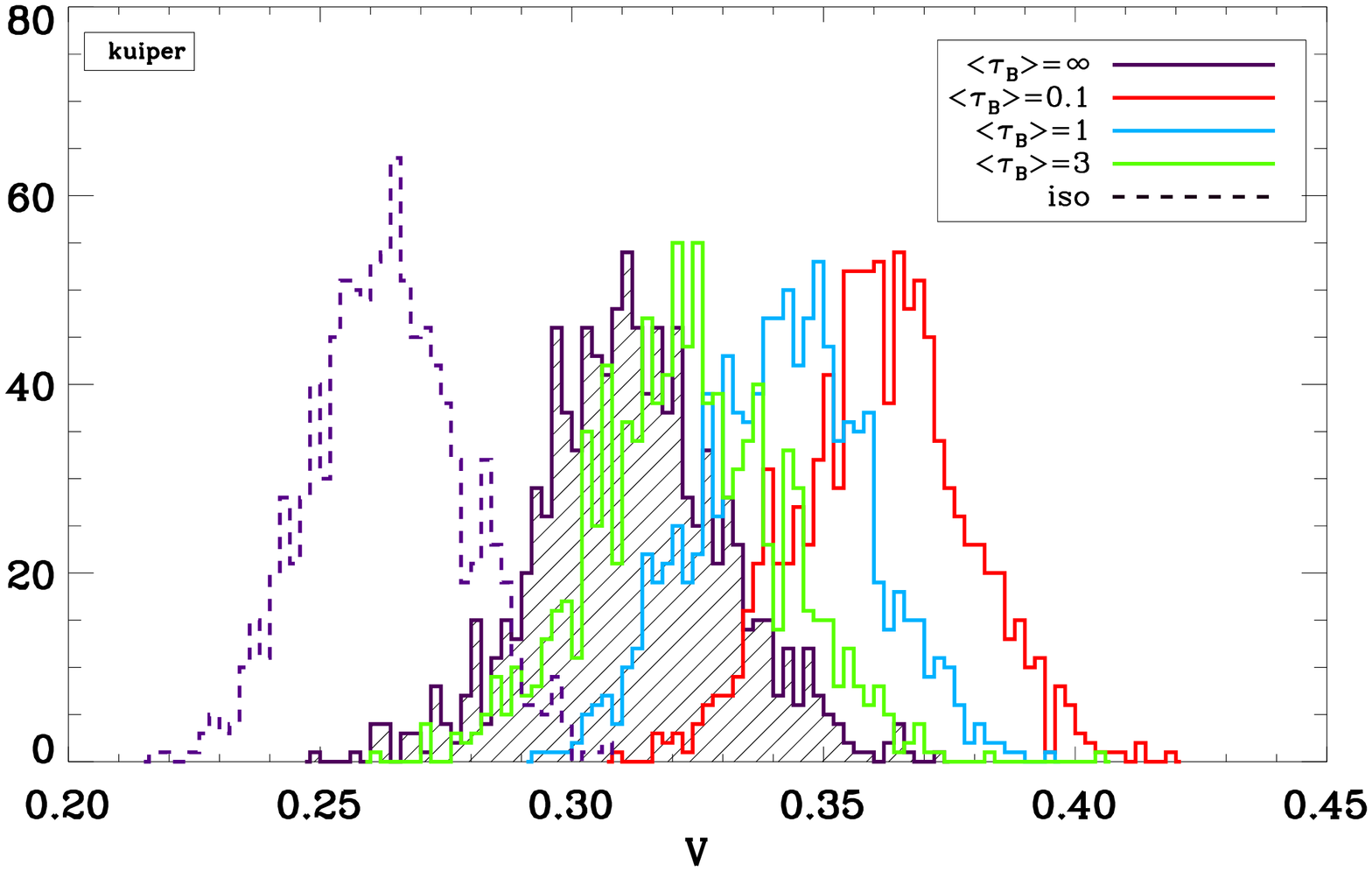}
\caption{Distribution of $V$ (top) and $D_{KS}$ (bottom) from 1000
  Monte Carlo realisation 1000 simulated events with energy above 60
  EeV. Mean source density $n_{\rm s} = 10^{-3}\rm Mpc^{-3}$. The dotted
  curves represent the isotropic model, the hatched curves denote the
  LSS model and from right to left we have $\langle\tau\rangle_{100}=
  0.1$, $\langle\tau\rangle_{100}= 1$, and $\langle\tau\rangle_{100}=
  3$.}

\label{fig:histo_kuiperDKS2d}
\end{figure}

\begin{figure}[th]
\centering

\includegraphics[angle = 90,width=9cm]{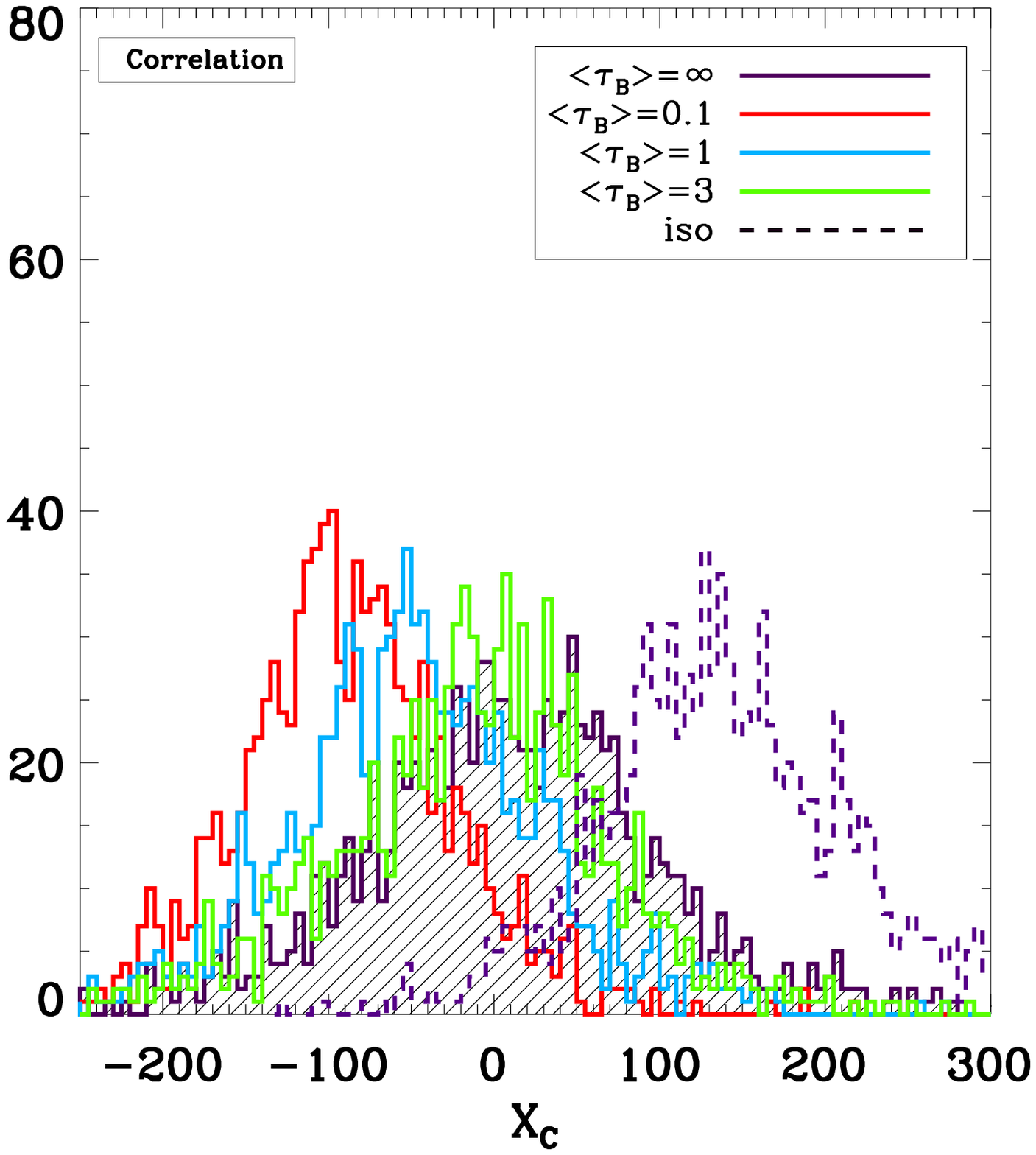}
\caption{Distribution of $X_{C}$ from 1000 Monte Carlo realisations of of
  1000 simulated events with energy above 60~EeV. The adopted mean
  source density is $n_{\rm s} = 10^{-5}\rm Mpc^{-3}$. The dotted
  curves represent the isotropic model, the hatched curves denote the
  LSS model and from left to right we have $\langle\tau\rangle_{100}=
  0.1$, $\langle\tau\rangle_{100}= 1$, and $\langle\tau\rangle_{100}=
  3$.}

\label{fig:histo_corNs5}
\end{figure}

With that being defined, we plot the histograms of the $X_{C}$, $V$,
and $D_{KS}$ distributions from 1000 Monte Carlo realizations. In
detail, Fig.~\ref{fig:histo_cor} presents the distributions of $X_{C}$
from 1000 Monte Carlo realisations of 100 simulated events of energy
above 80 EeV (top panel) and 1000 simulated events with energy above
60 EeV (bottom panel). In the top panel, we note some overlap between
the LSS and the other models with finite $\langle\tau\rangle_{100}$,
 the magnitude of which depends on the value of $\langle\tau\rangle_{100}$.
However, the effect of the finite magnetic optical depth becomes
pronounced in the lower panel, representing the case of $N_{\rm ev} = 1000$
above $60\,$EeV. We note that these statistics are for instance well within the capabilities of the JEM-EUSO
mission. In Fig.~\ref{fig:histo_kuiperDKS2d}, we plot the
distributions of $V$ (top) and $D_{KS}$ (bottom) for 1000 Monte Carlo
realisations of 1000 simulated events with energy above 60 EeV; we
have chosen this set of events for illustration because it appears
more sensitive, as indicated by the previous figure.  Qualitatively
speaking, the $X_{C}$ distribution appears to be a more efficient
discriminator of the effect discussed here. One might resort to definite
statistical tests to quantify the sensitivity of these tests
to the effect of finite optical depth but, given the exploratory
nature of this discussion, the general trend indicated by the above
figures suffices.

The comparison of the top and bottom panels of
Fig.~\ref{fig:histo_cor} indicates that a larger number of events is
more important in probing this effect, than a stronger anisotropy of the
arrival directions. 

Figure~\ref{fig:histo_corNs5} allows us to probe the sensitivity of the
above results to the mean source density. In this figure, we show the
distribution of $X_{C}$ from 1000 Monte Carlo realisations of 1000
simulated events of energy above 60~EeV for $n_{\rm s} = 10^{-5}\rm
Mpc^{-3}$, which corresponds to the existing lower bound on $n_{\rm
  s}$. When comparing with the bottom panel of figure~
\ref{fig:histo_cor} where $n_{\rm s} = 10^{-3}\rm Mpc^{-3}$, we note
that a lower source density implies a relatively weaker sensitivity to
the effect of finite magnetic optical depth. This results directly
from the stronger clustering of events in given locations of the sky,
which limits the number of directions with which one can sample the
sky maps shown in Fig.~\ref{fig:flux200}. Several remarks are in order
in this respect. First of all, one may hope that future large-scale
experiments will be able to determine the mean source number density
from the statistics of multiplets. If not, at the very least, one
would infer that $n_{\rm s}\gtrsim 10^{-3}\,$Mpc$^{-3}$, hence for the
present purpose the exact value of the present parameter is
unimportant. Assuming that the mean source density is known, it would
be interesting to explore the information contained in the
statistics of multiplets in different regions of the sky, as these
would obviously provide additional tools that are sensitive to the
above effect; this a study requires us to handle a larger dimensionality
of parameter space, thus is left to future work.

\subsection{Application to the Pierre Auger Observatory data}
To date, the only catalog of events above $60\,$EeV publicly available
is that released by the Pierre Auger Collaboration
\citep{Abraham2008b}. We thus now use this catalog as a
testbed for the previous tests. Since the catalog contains 27 events
above $57\,$EeV, which is reduced further to 18 by applying the PSCz
mask, the statistics is obviously limited and the present exercise
should be taken as an example. 

We use the Auger aperture given by
$$W(\delta) = \cos(a_0)\cos(\delta)\sin(\alpha_m) + \alpha_m
\sin(a_0)\sin(\delta),$$ where $a_0 = -35 ^{\circ}$ is the Auger
southern site latitude and $\alpha_m$ is given by
$$\alpha_m =
 \left\{
 \begin{array}{rl}
   0   &  \mbox{if }  \xi > 1  \\
   \pi &  \mbox{if }  \xi < -1 \\
   cos^{-1}(\xi) &    \mbox{ otherwise  } 
  \end{array}
  \right.
  $$  
  and $$\xi = \frac {\cos(\theta_{\rm max}) - \sin(a_0)\sin(\delta)}{\cos(a_0)\cos(\delta)}\,.$$
 
\begin{figure}[th]
\centering

\includegraphics[angle = 90,width=9cm]{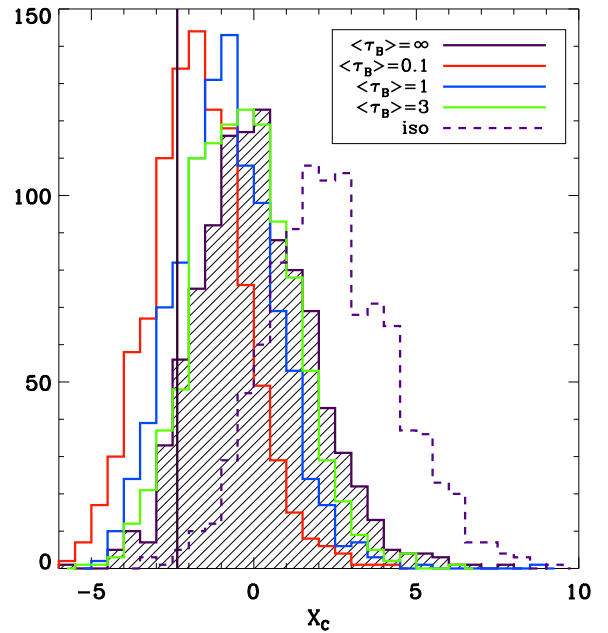}
\caption{Distribution of $X_{C}$ from 1000 Monte Carlo realizations of
  18 simulated events of energy above 60~EeV. The mean source
  density is $n_{\rm s} = 10^{-5}\rm Mpc^{-3}$. The dotted curves
  represent the isotropic model, the hatched curves denote the LSS
  model, and from left to right we have $\langle\tau\rangle_{100} =
  0.1$, $\langle\tau\rangle_{100} = 1$, and $\langle\tau\rangle_{100}
  = 3 $. The vertical line corresponds to the value obtained for the
  Pierre Auger data.}

\label{fig:histo_auger}
\end{figure}

  Figure~(\ref{fig:histo_auger}) shows the statistical distributions
  of $X_{C}$ reconstructed for the different models using Monte
  Carlo simulations, while the vertical line indicates the value
 obtained for the Auger events. In accordance with the choice of
  threshold energy, we have limited the horizon to 200~Mpc.  There is
  an almost complete overlap between the histograms of the LSS and the
  other models. Our study confirms that the statistics of these
  published Pierre Auger events appear inconsistent with the isotropic
  model and consistent with the LSS model (see
  \citealp{Kashti,Koers,Aublin}). We also recover a (marginal)
  tendency, as noted in \cite{Kashti}, for the arrival
  directions to favor a scenario in which the apparent arrival
  directions cluster more strongly than the LSS traced through the
  PSCz; in our case, this is represented by the $X_{C}$ test for the Auger events tending to measure smaller
  $\langle\tau\rangle_{100}$ values, although we
  note that this tendency remains at most marginal.

We noted in figures~\ref {fig:histo_cor} and
\ref{fig:histo_kuiperDKS2d} that the studied effect becomes
noticeable when the number of simulated events with energy above 60
EeV is of the order of $10^2-10^3$; as expected, the effect is too weak
to affect the Pierre Auger results and allow any clear distinction
between the different models.

\section{Conclusions}

We have discussed the possible distortion of the sky
maps of UHECR arrival directions associated with the inhomogeneous
distribution of large-scale magnetic fields.  As the effect of the
extragalactic magnetic field on the UHECR trajectories may be
described by a set of stochastic interactions with magnetized
structures, its influence can be conveniently characterized by the
optical depth to magnetic deflection $\tau$. Assuming that the UHECR
sources are of the bursting type, such as gamma-ray bursts or newly
born magnetars, and that their occurrence rate in the GZK
sphere is low, we have argued that the relative lack of magnetized structures
in those areas of the sky that have $\tau<1$ should lead to a depleted
cosmic-ray flux from these directions compared to those detected
from these directions if one assumed that the sources were steady
emitters. This is mainly related to the particles in those areas of
the sky, having a probability $\tau$ of interacting with a
magnetized structure, which in the above scenario represents a
necessary condition for the source to be observable in an experiment
lifetime. It is important to emphasize that the effect that we have
discussed here is different from the standard angular deflection
associated with extragalactic magnetic fields. While angular deflection
modifies the arrival direction without altering the flux, the effect
that we discuss directly affects the flux from some regions of the sky.

To quantify this distortion, we have constructed sky maps of
UHECR arrival directions from bursting sources residing in the large
scale structure for two energy thresholds of 60~EeV and 80~EeV (see
Figs.~\ref{fig:flux200} and \ref{fig:flux100}). We have considered
different configurations of the large-scale magnetic field and taken
into account the modulation associated with the probability of
experiencing at least one interaction with a magnetized system
$p_\tau=1-\exp(-\tau)$.  Each model is then characterized by the value
of $\langle\tau\rangle_{100}$, which expresses the average optical depth
of magnetic deflection calculated to a depth of 100 Mpc.  We have then
used various statistical tests previously described in the literature,
particularly the $X_{C}$ test [see Eq.~(\ref{eq:Xc})], which measures
the correlation between the simulated events from the model to be
tested and the average expected number of events from two reference
models, one model with isotropic arrival directions and one model in
which the sources are distributed according to the large-scale
structure such that the effect of finite magnetic optical depth is
neglected. Inspired by the statistics expected for future large-scale
experiments, we have applied these tests to two sets of events:
$N_{\rm ev}$ = 100 events with $E >$ 80 EeV and $N_{\rm ev}$ = 1000
events with $E > 60$~EeV, respectively.  The histograms from the second
set of events show more distinguishable models because of the high
number of expected events (see Figs.~\ref{fig:histo_cor} and
\ref{fig:histo_kuiperDKS2d}). We have also considered the catalog of
events (27 events above 57~EeV) released by the Pierre Auger
Observatory as an example.

In conclusion, we have found that the distortion examined here cannot affect
present scale experiments but should be considered when
performing anisotropy studies with future large-scale experiments. We
have noted a slight sensitivity of the distortion strength to the mean ultrahigh
energy cosmic ray source density, which would however disappear if
this latter were determined using the statistics of multiplets.

\bibliographystyle{aa}
\bibliography{KLK}
\end{document}